\documentclass{optica-article}

\journal{opticajournal} 

\articletype{Research Article}

\usepackage[separate-uncertainty=true]{siunitx}
\DeclareSIUnit\belm{Bm}
\usepackage{subfiles}
\usepackage{graphicx}
\usepackage{subfig}
\usepackage[capitalise]{cleveref}
\usepackage{ulem}

\begin{document}

\title{Near-unity quantum interference of transverse spatial modes in an ultra-compact inverse-designed photonic device}

\author{Jamika Ann Roque,\authormark{1} Daniel Peace,\authormark{2,3} Simon White,\authormark{4,5} Emanuele Polino,\authormark{4,5} Sayantan Das,\authormark{2,3,4} Farzard Ghafari,\authormark{4,5} Sergei Slussarenko,\authormark{4,5} Nora Tischler\authormark{4,5} and Jacquiline Romero\authormark{2,3,*}}

\address{\authormark{1}National Institute of Physics, College of Science, University of the Philippines, Diliman, Quezon City, 1101, Philippines\\
\authormark{2}Australian Research Council Centre of Excellence for Engineered Quantum Systems, University of Queensland, St Lucia, 4072, Australia\\
\authormark{3}School of Mathematics and Physics, University of Queensland, St Lucia, 4072, Australia\\
\authormark{4}Australian Research Centre of Excellence for Quantum Computation and Communication Technology, Griffith University, Nathan, QLD, 4111, Australia\\
\authormark{5}Centre for Quantum Dynamics, Griffith University, Nathan, QLD, 4111, Australia}
\email{\authormark{*}m.romero@uq.edu.au} 


\begin{abstract*} 
The transverse spatial mode of photons is an untapped resource for scaling up integrated photonic quantum computing. 
To be practically useful for improving scalability, reliable and high-visibility quantum interference between transverse spatial modes on-chip needs to be demonstrated. 
We show repeatable quantum interference using inverse-designed transverse mode beamsplitters that have an ultra-compact footprint of \qty{3} {\micro\metre} $\times$ \qty{3} {\micro\metre}---the smallest transverse mode beamsplitters for \qty{1550} {\nano\metre} photons to date. We measure a Hong-Ou-Mandel visibility of up to 99.56$\pm$0.64 \% from a single device, with an average visibility across three identical devices of 99.38$\pm$0.41 \%, indicating a high degree of reproducibility. Our work demonstrates that inverse-designed components are suitable for engineering quantum interference on-chip of multimode devices, paving the way for future compact integrated quantum photonic devices that exploit the transverse spatial mode of photons for high-dimensional quantum information.
\end{abstract*}


\section{Introduction}

%

Quantum technologies have the potential to transform communication, computation, and sensing \cite{sood2023quantum, sood2024archives, pelucchi2022potential}. In many of these applications photons play a significant role: As the quantum system for quantum information processing, as a carrier of quantum information between spatially separated quantum nodes, or as a necessary ingredient for other physical platforms (e.g. excitation for trapped ions). As a quantum system, photons maintain coherence at room temperature and over long times, making photons desirable for many quantum applications\cite{slussarenko2019photonic}. In particular, photonic-based architectures are strong contenders for future scalable quantum computers \cite{psiquantum2025manufacturable, xanadu2025scaling}. The scalability, complexity, and energy efficiency required for such systems is often not possible with traditional table-top bulk optics experiments \cite{lu2021advances, pelucchi2022potential}. The role of integration for quantum photonic devices is increasingly recognised as a necessary step towards having practically relevant quantum technologies, reflected in the growing field of quantum photonic integrated circuits (QPICs). 

To date, the majority of QPICs have primarily utilised path, polarisation, time-bin, and frequency as degrees of freedom (DOF) to carry quantum information \cite{bao2023verylargescale, wan2024multidimensional, montaut2025progress, myilswamy2024on-chip, corrielli2021femtosecond}. Similarly, the transverse spatial mode of the photons has typically been underutilised in classical applications, though it has gained attention in recent years \cite{li2018multimode, cristiani2022roadmap}. Whether in the quantum or classical regime, the use of a variety of DOFs is beneficial to combat the ever-increasing information volume and Hilbert space requirements. In classical optical interconnects, the ability to co-propagate multiple wavelengths and transverse modes in a single physical channel has been used to densely encode information, thereby increasing information capacity and shoreline density \cite{yang2022multi,wan2024multidimensional,sun2025edge}. While multimode devices (we will use \textit{multimode} to mean multiple transverse modes, for brevity) can present additional design challenges, a variety of components required for both classical and quantum applications have been developed in recent years, examples include mode multiplexers \cite{jiang2019compact}, multimode waveguide crossers \cite{xu2018metamaterial2mode, guo2023transmitarray3mode}, bends \cite{jiang2018low-loss}, and phase shifters \cite{priti2019reconfigurable}. 

Our work focuses on the use of transverse modes in the quantum regime. The coherent conversion and control between path, polarisation, and transverse mode DOFs\cite{feng2016chip}, along with the generation of transverse-mode-entangled photon pairs on chip \cite{feng2019chip}, have both been recently demonstrated. These capabilities open up new possibilities of simultaneously utilising and controlling these DOFs to increase dimensionality in quantum information processing with integrated photonics. The observation of Hong-Ou-Mandel (HOM) interference \cite{hong1987measurement} between transverse waveguide modes, facilitated by a Bragg grating acting as a mode beamsplitter, further underscores the potential of transverse modes in complex quantum operations \cite{mohanty2017quantum}. More recently, the implementation of a linear optical controlled-NOT (CNOT) gate that acts on transverse waveguide modes has been demonstrated using two-mode directional couplers and multimode attenuators \cite{feng2022transverse}. 

Despite this progress, large-scale implementations that use transverse modes have yet to be realised---in part constrained by the arduous nature of traditional multimode component design methods. Multimode devices are usually larger and typically designed via a bottom-up approach, wherein devices are built from known first principles. Bottom-up approaches often lead to designs that are sensitive to fabrication error and limited in functionality. For example in Ref.~\cite{mohanty2017quantum}, the grating structure based on a first principles approach only facilitates HOM interference between modes of a given parity depending on the chosen grating symmetry. 

Another approach to designing components is to start with the function of the device, as in inverse design \cite{Su2020}. In contrast to the bottom-up approach, inverse design specifies the desired performance. Iterative brute force optimisation techniques are used to reach the desired performance, powered by the increase in computing power which allows significantly larger parameter spaces to be explored. In particular, gradient based optimisation techniques which utilise the adjoint method ensure that the gradient of each parameter within the space, regardless of size, is calculable by performing only two simulations (forward and adjoint) \cite{Lalau-Keraly2013}. As a result, topological inverse design has become a leading approach for designing many photonic devices due to the vast parameter space, enabling devices with compact footprint \cite{Callewaert2016}, broadband operation \cite{hansen2024inverse} and improved fabrication robustness \cite{shang2023inverse-designed}.

In the classical regime, the use of inverse design has already been shown to markedly improve performance and decrease the dimensions of multimode devices for mode multiplexing \cite{jiang2023inverse, zhang2021scalable, Shen2024} and bends \cite{yang2022inverse}. Inverse design has also been used for ultra-compact quantum CNOT and Hadamard gates in path-encoded photonic qubits \cite{he2023super}. More recently, multi-photon quantum interference has been demonstrated using an inverse-designed tritter \cite{huang2025multiphoton}. In this work, we use inverse design to create ultra-compact mode beamsplitters and mode multiplexers on-chip. Combining these components, we demonstrate near-unity Hong-Ou-Mandel interference between transverse waveguide modes. Our results show the suitability of inverse-designed components for high-performance quantum information processing that use transverse modes.

\section{Hong-Ou-Mandel interference in lossy, unbalanced, asymmetric beamsplitters} \label{sec:HOMtheory}

\begin{figure}
    \centering
    \includegraphics[width=\linewidth, page=1]{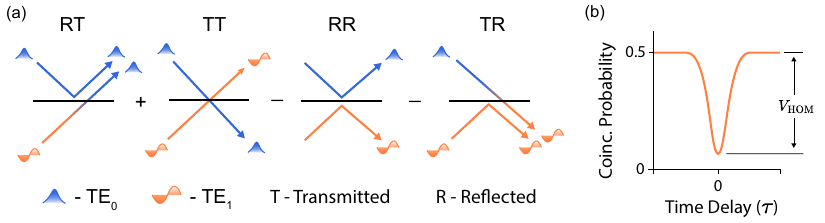}
    \caption{(a) Four possible combinations for two photons incident on a beamsplitter. If the incident photons are indistinguishable, then scenarios in which both photons are transmitted (TT) or reflected (RR), cannot be distinguished and therefore their probability amplitudes cancel due to opposite phases. As a result only cases where the photons bunch in one of the outputs (RT or TR) are observed. (b) Experimentally, the interference is observed as a dip in the coincidences when the time delay $\tau$ between the two photons approach zero.}
    \label{fig:hom_illustration}
\end{figure}

Hong-Ou-Mandel interference is a quintessential photonic quantum phenomenon \cite{hong1987measurement, bouchard2020two}. Unlike other photonic experiments that can be explained by classical or semi-classical models, understanding the HOM interference requires quantisation of the electromagnetic field. A Hong-Ou-Mandel experiment consists of two photons, each one incident on different input ports of a beamsplitter. If the two photons are indistinguishable, the probability amplitudes of the cases where the photons exit the beamsplitter from different output ports (i.e. cases where both are transmitted or both reflected, as illustrated in \cref{fig:hom_illustration}(a)) destructively interfere. Quantum interference leads to the quintessential \textit{bunching} effect, where both photons tend to exit the beamsplitter through the same output port. This bunching results in the characteristic HOM dip shown in \cref{fig:hom_illustration}(b). As the time delay $\tau$ between the two photons approach zero (i.e. the photons become indistinguishable), the photons exit the same port and the number of coincidences drop. The visibility of the HOM dip ($V_{\text{HOM}}$ in  \cref{fig:hom_illustration}(b)) depends on the indistinguishability of the two photons and the properties of the beamsplitter, in particular its symmetry. In the ideal case, the beamsplitter is assumed to be lossless with a balanced and symmetric splitting ratio between the outputs. In practice, all beamsplitters have some degree of loss and imperfect splitting ratio. 

In general the relationship between the input and output fields incident on a beamsplitter is given by the transfer matrix,
\begin{equation}
\begin{pmatrix}
E_3 \\
E_4
\end{pmatrix} = 
\begin{pmatrix}
t_1 & r_2e^{i\theta_2} \\ 
r_1e^{i\theta_1} & t_2e^{i\theta_3}
\end{pmatrix}
\begin{pmatrix}
E_1\\ 
E_2
\end{pmatrix}
\label{eq:generalBS}
\end{equation}

where $t_i$ and $r_i$ are the real amplitudes of the transmission and reflection coefficients for mode $i$ of the beamsplitter. In this form, the beamsplitter can be lossy ($t_i^2+r_i^2<1$), unbalanced ($t_i \neq r_i$), and asymmetric ($t_1\neq t_2, r_1\neq r_2$). Beamsplitters are typically encountered as operating on orthogonal paths. However, they can also be designed to operate on other orthogonal modes, such as transverse spatial modes. In this work, we consider spatial mode beamsplitters where an input light of either waveguide modes $\text{TE}_0$ or $\text{TE}_1$ from a multimode waveguide is split between $\text{TE}_0$ and $\text{TE}_1$. Furthermore, instead of the more common approach of only considering two phase terms, we also consider a third phase term $\theta_3$ when writing down the beamsplitter matrix. This is especially relevant in our case of spatial mode beamsplitters, since the transmitted light from each mode can have non-zero phase differences. In a lossless  beamsplitter, energy conservation dictates that the phases follow the condition 
\begin{equation}
    \alpha \equiv \theta_1+\theta_2-\theta_3 = \pi
\end{equation}\label{eq:phasecondition} 
and that  the coefficient amplitudes should be symmetric ($t_1=t_2, r_1=r_2$).

In this work, we perform the HOM experiment by measuring the coincidences as a time delay $\tau$ is applied on one of the arms. The coincidence probability as a function of $\tau$ is given by \cite{fedrizzi2009anti}, 
\begin{equation}\label{eq:star}
P(1,1)^{\tau}_{1,2}=r_1^2r_2^2+t_1^2t_2^2+2r_1r_2t_1t_2\cos{\alpha}I_{\text{overlap}}(\tau)
\end{equation}
where $I_{\text{overlap}}$ is the overlap integral of the joint spectral amplitude of the two photons. The dependence of the HOM interference on the properties of the beamsplitter can be described by defining an effective splitting ratio $\eta_{\text{eff}}$ as,
\begin{equation}
\label{eq:neff}
\eta_{\text{eff}} = \frac{t_0^2}{r_0^2+t_0^2},
\end{equation}
where $r_0=\sqrt{r_1r_2}$ and $t_0=\sqrt{t_1t_2}$. Therefore, the expected HOM interference visibility based on the properties of the beamsplitter and photon pair joint spectral amplitude can be expressed as
\begin{equation}\label{eq:vhomneff}
    V_{HOM} = -\frac{2  \eta_{\text{eff}}(1-\eta_{\text{eff}})\cos{\alpha}I_{\text{overlap}}(0)}{(1-2\eta_{\text{eff}}+2\eta_{\text{eff}}^2)}.
\end{equation}

For additional details and full derivation see Section 8 of Supplement 1.

\begin{figure}
    \centering
    \includegraphics[width=\linewidth, page=2]{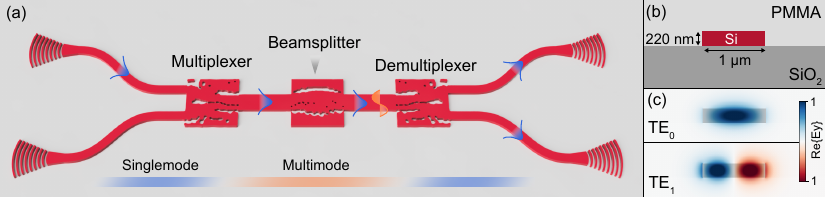}
    \caption{(a) Schematic diagram of the photonic chip for measuring quantum interference between transverse modes in a multimode waveguide. TE$_0$ mode in a single-mode waveguide propagates via a multiplexer to a multimode waveguide that is input to a transverse mode beamsplitter. After mode-mixing at the beamsplitter, the TE$_0$ and TE$_1$ modes are routed back to single-mode waveguides via a demultiplexer.  (b) Multimode waveguide cross-section showing the {\qty{1} {\micro\metre} x \qty{220} {\nano\metre}} silicon-on-insulator waveguide with a PMMA cladding. (c) Electric field profiles for the first two transverse electric (TE) modes of the multimode waveguide.} 
    \label{fig:chip-schematic}
\end{figure}

\section{Device Design}
The schematic of the device used for measuring HOM interference between transverse spatial modes is presented in \cref{fig:chip-schematic}(a). We used a standard silicon-on-insulator (SOI) wafer featuring a \qty{220} {\nano\metre} thick silicon device layer on top of a \qty{3} {\micro\metre} buried oxide (SiO$_2$) layer with a polymethyl methacrylate (PMMA) overcladding (shown in \cref{fig:chip-schematic}(b)). The chip consists of single-mode and multimode regions, with the multimode region supporting both TE$_0$ and TE$_1$ modes. Photons are coupled via grating couplers that are connected to single-mode waveguides (\qty{500} {\nano\metre} in width). The photons are routed to the  input multimode waveguide (\qty{1} {\micro\metre} in width) via an inverse-designed two-mode multiplexer.  Mode-mixing at the inverse-designed transverse mode beamsplitter results to TE$_0$ and TE$_1$ modes at the output multimode waveguide of the beamsplitter (\cref{fig:chip-schematic}(c)).  A demultiplexer (of the same design as the multiplexer) is placed after the output multimode waveguide to route the photons back into single-mode waveguides, before being coupled off-chip via the grating couplers. Both the beamsplitter and multiplexer are designed using the Stanford Photonic INverse design Software (SPINS), which performs a gradient-based optimisation using the adjoint method to efficiently calculate the gradient of the parameters which define the permittivity distribution $\epsilon$ (i.e.,~device topology) \cite{Su2020, hughes2018adjoint}. For each eigenmode source of a given wavelength and mode, the device response (forward) and its adjoint are simulated using the SPINS Finite-Difference Frequency-Domain (FDFD) solver. In general, the optimisation routine seeks to find the parameters, $p$ of the permittivity distribution $\epsilon(p)$ that minimises the objective function given by
\begin{equation}
    f(\epsilon)=\sum_\lambda\sum_m\sum_s(t_{ms}-|S_{ms}(\epsilon,\lambda)|)^2,
\end{equation}

where $t_{ms}$ is the target transmission for mode source $s$ (akin to an input port $s$) going through the mode overlap monitor $m$ (akin to an output port $m$). $S_{ms}$ denotes the device S-parameters, corresponding to the simulated transmissions of the respective input-output combinations (i.e. elements of the transfer matrix). Note that while SPINS can optimise for specific phases at each of the monitors \cite{Su2020}, we do not explicitly set any target phase in the objective function. However, the optimised beamsplitter designs demonstrate relative phase differences between the output modes that are consistent with energy conservation, as outlined in \cref{eq:phasecondition}, and as also previously observed \cite{Nanda2024ExploringOptimization}. Fabrication constraints such as minimum feature sizes are imposed following a levelset method introduced in the later stages of the optimisation routine \cite{Vercruysse2019}.

\begin{figure}
    \centering
    \includegraphics[page=3]{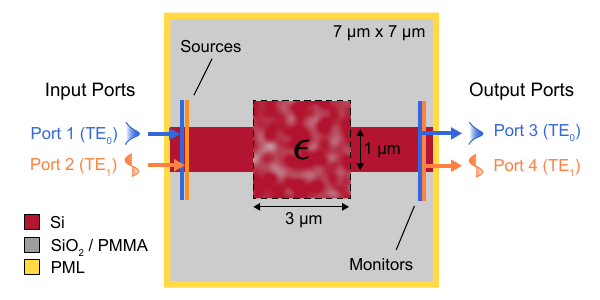}
    \caption{Simulation setup for the inverse-designed transverse mode beamsplitter. Mode sources (Ports 1 and 2) excite the $\text{TE}_0$ and $\text{TE}_1$ modes of the multimode waveguide, respectively. After passing through the \qty{3} {\micro\metre} $\times$ \qty{3} {\micro\metre} design area with permittivity $\epsilon$, the power coupled to each modes in the output waveguide is measured via the mode monitors (Ports 3 and 4). Simulation region is \qty{7} {\micro\metre} $\times$ \qty{7} {\micro\metre} $\times$ \qty{2} {\micro\metre} with perfectly matched layer (PML) boundary conditions.}
    \label{fig:mbs_setup}
\end{figure}

\Cref{fig:mbs_setup} illustrates the simulation setup for the mode beamsplitter optimisation in SPINS. We set the design area to \qty{3} {\micro\metre} $\times$ \qty{3} {\micro\metre}. The total simulation region was {\qty{7} {\micro\metre} $\times$ \qty{7} {\micro\metre}} $\times$ \qty{2} {\micro\metre} with a uniform mesh of pixel size \qty{44} {\nano\metre}, and perfectly matched layer (PML) boundary conditions. Optimisation of the beamsplitter was performed using the following objective function with the target transmission $t$ of each output mode set to {$\sqrt{0.5}$} for both mode sources:
\begin{equation}\label{eq:mbs_obj}
    f_{\text{mbs}}(\epsilon)=\sum_\lambda\sum_m^{3,4}\sum_s^{1,2}({\sqrt{0.5}}-|S_{ms}(\epsilon,\lambda)|)^2.
\end{equation}

\begin{figure}
    \centering
    \includegraphics[width=4.5in, page=4]{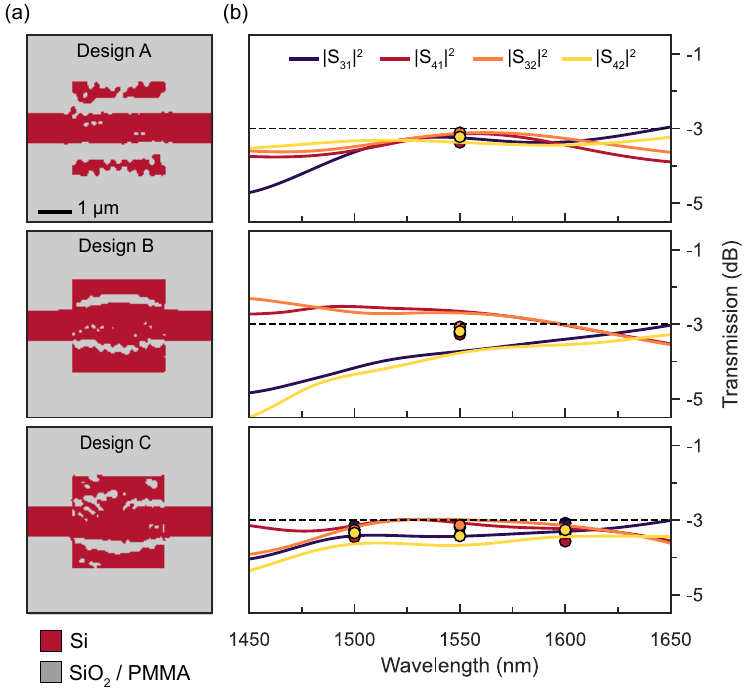}
    \caption{Inverse-designed transverse mode beamsplitters. (a) Final design of each optimised device. (b) Broadband simulated frequency response for each device. The solid coloured lines indicate the frequency response at the optimisation wavelength indicated by the corresponding coloured dots, the dashed lines indicate the ideal target frequency response.}
    \label{fig:mbs_sims}
\end{figure}

We explore several optimisation conditions resulting in three different devices we label as A, B, and C as shown \cref{fig:mbs_sims}(a). The first device (Design A) was optimised for a single wavelength ($\lambda = $\qty{1550} {\nano\metre}) and the minimum feature size constrained to \qty{120} {\nano\metre}, with the optimisation running for a total of 120 iterations. In subsequent optimisations (resulting to Designs B and C) the minimum feature size is reduced to \qty{80} {\nano\metre}, and the total number of iterations is reduced to 80. For Design B, we maintain the optimisation at a single wavelength of $\lambda = $\qty{1550} {\nano\metre}, while Design C is optimised across three wavelengths $\lambda = $\qty{1500} {\nano\metre}, \qty{1550} {\nano\metre}, and \qty{1600} {\nano\metre}. Given that the simulation dimensions and pixel size of the mesh is the same for all three devices, the optimisation times for each device scales linearly with the number of optimisation wavelengths and iterations. The total optimisation times are 7.5 hours, 5 hours and 15 hours for designs A, B and C, respectively. 

Using the same design area, simulation size, and pixel size as that of the optimisation for the mode beamsplitter, the mode multiplexer optimisation was performed on a single wavelength of $\lambda = $\qty{1550} {\nano\metre}, and minimum feature size of \qty{120} {\nano\metre} across a total of 100 iterations. In this case, the objective function is modified such that the first two terms maximise transmission into the desired mode while the last two terms are included to further reduce unwanted coupling to the other mode (crosstalk),
\begin{equation}\label{eq:mdm_obj}
\begin{split}
f_{\text{mdm}}(\epsilon)= (1-|S_{31}(\epsilon)|)^2+(1-|S_{41}(\epsilon)|)^2 \\
+|S_{32}(\epsilon)|^2+|S_{42}(\epsilon)|^2.
\end{split}
\end{equation}

More details about the optimisation procedure is outlined in Section 9.3 of Supplement 1.

\begin{figure}
    \centering
    \includegraphics[width=1\linewidth, page=5]{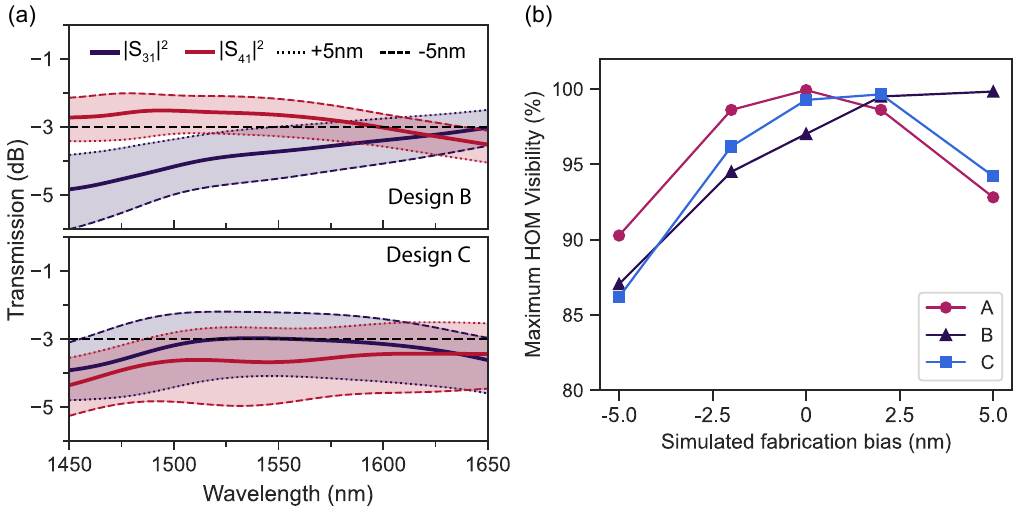}
    \caption{Simulated fabrication robustness. (a) Simulated $S_{31}$ (TE$_0$ to TE$_0$) and $S_{41}$ (TE$_0$ to TE$_1$) S-parameters (solid line) for Designs B and C. The shaded regions (bounded by the coloured dashed and dotted lines) denote a  \qty{\pm5} {\nano\metre} fabrication bias. (b) Maximum theoretical HOM visibilities versus fabrication bias based on \cref{eq:vhomneff} and the simulated effective splitting ratios.}
    \label{fig:bias}
\end{figure}

Following optimisation with SPINS, the broadband frequency response of each device is verified using a Finite-Difference Time-Domain (FDTD) (Flexcompute Inc, Tidy3D) solver with a high-resolution, non-uniform mesh. Results of the broadband simulations are shown in \cref{fig:mbs_sims}(b), with the solid coloured lines giving the frequency response of the device optimised for the wavelengths indicated by the  corresponding coloured dots.  Each device presents unique spectral characteristics. In particular, the response of Design B resembles that of a traditional directional coupler featuring a single crossing point at which the splitting is highly balanced and symmetric. (We note that the crossing point in Design B is offset from the optimisation wavelength, this is attributed to a mismatch in the simulation grids from the optimisation and from the FDTD.) In contrast, the response of A and C is generally flatter across the frequency range and can feature multiple crossing points. These characteristics are particularly evident in simulations of the fabrication tolerance of each device, shown in \cref{fig:bias}(a).  Here we apply a uniform bias to the device geometry to emulate the underetching or overetching in fabrication. For Design B, the crossing point shifts in wavelength towards shorter wavelengths for a positive bias (i.e. underetching). The flat responses of A and C lead to increased asymmetry across all wavelengths uniformly, with both a positive or negative fabrication bias increasing the asymmetry. Based on the simulated transmission and reflection coefficients, the maximum HOM visibilities when bias is applied are calculated following \cref{eq:vhomneff} and presented in \cref{fig:bias}(b).

\section{Experiment and Results}

\begin{figure}[htbp]
    \centering
    \includegraphics[width=1\linewidth, page=12]{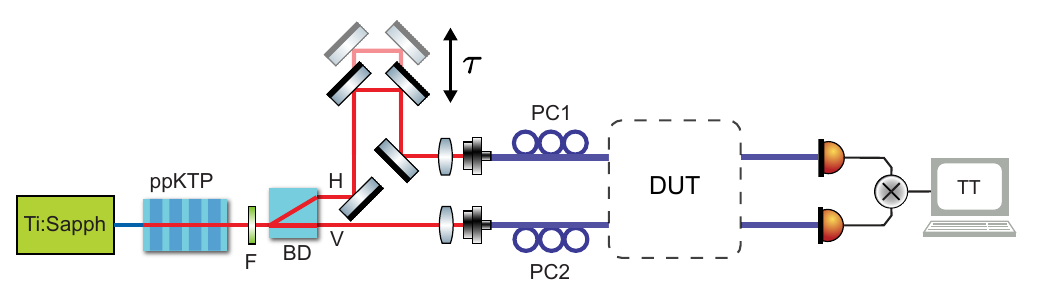}
    \caption{Simplified experimental setup for the HOM interference measurements. Degenerate photon pairs are produced off chip via type-II SPDC from a ppKTP crystal. Photons are spatially separated based on their polarisation (horizontal-H, vertical-V) after passing through the beam displacer (BD). For one of the photons, an adjustable path delay is used to add a time delay $\tau$ before coupling to the single-mode fibre. Single-mode fibres route the photons to and from the device under test (DUT) via the on-chip grating couplers. Coincidence measurements are recorded between the two outputs of the DUT using superconducting nanowire single-photon detectors (SNSPDs) and counting logic.}
    \label{fig:experimental_dut}
\end{figure}

We generate pairs of degenerate \qty{1548} {\nano\metre} photons outside the chip via type-II spontaneous parametric down-conversion (SPDC) in a bulk ppKTP crystal. The two photons are then separated based on their polarisation and each one is coupled into single-mode fibres\cite{ghafari2025entanglement, White2025timebin}. In one of the paths, an adjustable free-space delay is used to add a time delay $\tau$ to the arrival of one photon at the input of the device under test (DUT). The HOM interference is probed as a function of the relative delay between the two indistinguishable photons. The indistinguishability of the photons from the source is ascertained independently using a HOM interference setup outside the chip, for which we measure the source visibility to be \num{99.57 \pm 0.30}\unit{\percent} (see Section 10 of Supplement 1).

The single photons are coupled into and out of the chip via shallow-etched uniform grating couplers and a multichannel polarisation maintaining fibre array (Oz Optics). The coupling efficiency of the grating couplers is measured to be \qty{-5.46} {\decibel} at the wavelength of {\qty{1548} {\nano\metre}} (see Section 9.2 of Supplement 1), resulting in a total coupling loss of \qty{-10.92} {\decibel}. On chip, the multiplexers and beamsplitter combined contribute an additional \qty{-1} {\decibel} loss resulting in a total device efficiency following the source setup of \qty{-11.92} {\decibel}.

For each device, the coincidence counts (CC) were recorded for a total of 109 data points across a \qty{10} {\pico\second} time delay window ($\Delta\tau$). The scans were done with \qty{0.1} {\pico\second} $\tau$ increments, except for 13 points around $\tau=0$, where the increments were reduced to \qty{0.033} {\pico\second} in order to improve the quality of the fitting and resolve the lowest possible point of the dip. For each device we combine three separate coincidence measurements each \qty{40} {\second} long with a coincidence window of \qty{2} {\nano\second}, for a total integration time of 120 seconds per data point equivalent to approximately 3500 coincidences per bin when outside the dip.

\begin{figure}
    \centering
    \includegraphics[width=1\linewidth, page=6]{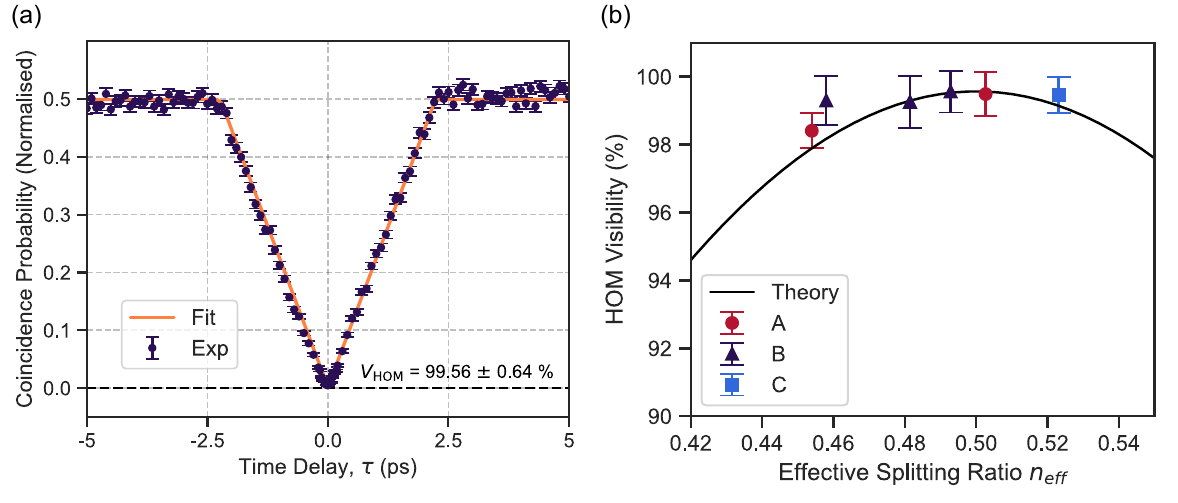}
    \caption{Measured on-chip HOM visibilities. (a) Normalised coincidences versus relative path delay $\tau$ for the best performing device with a measured visibility of $V_{\text{HOM}} =$ 99.56$\pm$0.64 \%. Markers are experimental data points with error bars representing standard deviation assuming Poissonian statistics. Orange line indicates triangular fit based on Ref.~\cite{fedrizzi2009anti} with error given as standard deviation. (b) Experimentally measured HOM visibilities for each beamsplitter design. Solid line indicates the theoretical HOM visibility based on the measured effective splitting ratio of each device as given by \cref{eq:vhomneff}. Data points represent experimentally measured HOM visibilities.}
    \label{fig:HOM_results}
\end{figure}

From our best performing device featuring beamsplitter Design B, with $\eta_{\text{eff}} =$ \qty{48.77} {\percent} (based on classical characterisation; see Section 9.3 of Supplement 1) we measure a visibility of $V_{\text{HOM}} =$ 99.56$\pm$0.64 \% shown \cref{fig:HOM_results}(a). To confirm the reproducibility of the inverse-designed devices, we measure visibility for multiple identical copies fabricated on the same chip. For the best performing device which features Design B, the average measured visibility across three devices is 99.38$\pm$0.41 \% (shown \cref{fig:HOM_results}(b)). Similarly, we measure multiple devices featuring Design A and C, with corresponding visibilities shown in \cref{fig:HOM_results}(b). The measured visibilities are plotted against the ones theoretically predicted by \cref{eq:vhomneff} based on $\eta_{\text{eff}}$ values derived from classical characterisation. All three inverse-designed beamsplitters lead to visibilities that are higher than 98\%.

\section{Discussion}

In this work, we demonstrate the usefulness of inverse-designed elements for multimode quantum photonic devices. We demonstrate high (above 98\%) interference visibilities using ultra-compact devices, with values matching the ones predicted using the model in \cref{eq:vhomneff}. Thus, with a perfect SPDC source, we expect to achieve visibilities of up to \qty{99.95} {\percent} with our best device. If the mode beamsplitter is designed based on nanoscale gratings \cite{mohanty2017quantum}, we estimate the beamsplitter length to be \qty{60} {\micro\metre} (see Section 13 of Supplement 1). The inverse design approach presented in this work offers a factor of 20 reduction in the size of the beamsplitter. The small features sizes (\qty{10} {\nano\metre}) required for the design based on nanoscale gratings greatly limit its fabrication potential. On the other hand, the feature size can be constrained in the inverse design approach, leading to improved fabrication robustness.


With the inverse design method, the splitting ratios can be easily customised without having to vary the device dimensions. The beamsplitters can also be easily scaled up to more modes, as we demonstrate by designing a three-port beamsplitter (tritter) using the same design area, and with minimal changes to the optimisation procedure (details in Section 12 of Supplementary 1). 

Importantly, we show that inverse-designed elements are repeatable. Performance of duplicate copies of each design shows consistently good performance. By simulating overetched and underetched conditions in the mode beamsplitter designs, we showed that theoretical visibilities of at least \qty{85} {\percent} are maintained within a fabrication bias range of \qty{\pm5} {\nano\metre} (above \qty{90} {\percent} for \qty{\pm2.5} {\nano\metre} bias and above the classical threshold of \qty{50} {\percent} for \qty{\pm20} {\nano\metre}). Because HOM visibility is only dependent on the effective splitting ratio of a device, a lossy device can still maintain good visibilities as long as the losses are balanced. Fabrication robustness can further be enhanced by incorporating device bias within the optimisation routine itself \cite{shang2023inverse-designed}. 

The qudit measurement space is immense, consisting of arbitrary transformations on single or many qudits. High-dimensionality is a double-edged sword---while the potential of the enlarged Hilbert space brings benefits \cite{wang2020qudits, kiktenko2023realization}, it also brings challenges in terms of the complexity of devices for manipulating quantum information. An elegant approach to decomposing qudit unitary transformations recursively into smaller unitary operations has been developed in \cite{de2018simple}. Another method based on multimode blocks rather than the usual two-mode blocks has been proposed in \cite{saygin2020robust}. While these methods do not necessarily improve scaling over more commonly used Reck-Zeilinger \cite{reck1994experimental} and Clements \cite{clements2016optimal}, these recent approaches  based on multimode blocks are more amenable to experimental parametrisation. We expect that photonic inverse design will be important in designing these units for quantum information processing in the future.

\section{Conclusion}
This work highlights the use of photonic inverse design to overcome the complexity of designing multimode components. We show that inverse design leads to devices of compact form factor that can be used in the quantum regime.  We show that despite their irregularly shaped features, inverse-designed components are reproducible and can exhibit consistently high-visibility quantum interference. This is particularly important for realising the potential of transverse modes and other high-dimensional photonic degrees of freedom for quantum information processing. In particular, the components in this work can be combined with transverse mode entangled photon sources \cite{feng2019chip} and mode-dependent phase shifters \cite{priti2019reconfigurable} to enable a new approach to programmable multimode quantum photonics devices.



\begin{backmatter}

\bmsection{Funding}
J.A.R.\ was supported by the Department of Science and Technology-Accelerated Science and Technology Human Resource Development Program (DOST-ASTHRDP) through the Graduate Studies Scholarship and the Research Enrichment Program grant. This work was supported by the Australian Research Council (ARC) Centre of Excellence for Engineered Quantum Systems (EQUS, CE170100009) and Centre of Excellence for Quantum Computation and Communication Technologies (CQC2T, CE170100012). N.T. was supported by an ARC Discovery Early Career Researcher Award (DE220101082). F.G\ was supported by the Griffith University Postdoctoral Fellowship (GUPF58938).

\bmsection{Acknowledgments}
\noindent The authors acknowledge the use of the High Performance Computing facilities of The University of the Philippines - Data Commons. The authors also acknowledge the use of NCRIS-enabled facilities including both Microscopy Australia at the Centre for Microscopy and Microanalysis (CMM), the University of Queensland and the Queensland node of the Australian National Fabrication Facility (ANFF).
While preparing this manuscript, we became aware of a related work \cite{huang2025multiphoton}, that was recently posted on the arXiv.
\bmsection{Disclosures}
\noindent The authors declare no conflicts of interest.

\bmsection{Data availability} 
\noindent Data underlying the results presented in this paper are available from the corresponding authors upon reasonable request.

\bmsection{Supplemental document}
See Supplement 1 for supporting content. 

\end{backmatter}


\begin{thebibliography}{10}
\newcommand{\enquote}[1]{``#1''}

\bibitem{sood2023quantum}
S.~K. Sood \emph{et~al.}, \enquote{Quantum computing review: A decade of research,} {\protect\JournalTitle{IEEE Transactions on Engineering Management}} \textbf{71}, 6662--6676 (2023).

\bibitem{sood2024archives}
V.~Sood and R.~P. Chauhan, \enquote{Archives of quantum computing: research progress and challenges,} {\protect\JournalTitle{Archives of Computational Methods in Engineering}} \textbf{31}, 73--91 (2024).

\bibitem{pelucchi2022potential}
E.~Pelucchi, G.~Fagas, I.~Aharonovich, \emph{et~al.}, \enquote{The potential and global outlook of integrated photonics for quantum technologies,} {\protect\JournalTitle{Nature Reviews Physics}} \textbf{4}, 194--208 (2022).

\bibitem{slussarenko2019photonic}
S.~Slussarenko and G.~J. Pryde, \enquote{Photonic quantum information processing: A concise review,} {\protect\JournalTitle{Applied Physics Reviews}} \textbf{6} (2019).

\bibitem{psiquantum2025manufacturable}
P.~Team, \enquote{A manufacturable platform for photonic quantum computing,} {\protect\JournalTitle{Nature}} pp. 1--3 (2025).

\bibitem{xanadu2025scaling}
H.~Aghaee~Rad, T.~Ainsworth, R.~Alexander, \emph{et~al.}, \enquote{Scaling and networking a modular photonic quantum computer,} {\protect\JournalTitle{Nature}} pp. 1--8 (2025).

\bibitem{lu2021advances}
L.~Lu, X.~Zheng, Y.~Lu, \emph{et~al.}, \enquote{Advances in chip-scale quantum photonic technologies,} {\protect\JournalTitle{Advanced Quantum Technologies}} \textbf{4}, 2100068 (2021).

\bibitem{bao2023verylargescale}
J.~Bao, Z.~Fu, T.~Pramanik, \emph{et~al.}, \enquote{Very-large-scale integrated quantum graph photonics,} {\protect\JournalTitle{Nature Photonics 2023 17:7}} \textbf{17}, 573--581 (2023).

\bibitem{wan2024multidimensional}
Y.~Wan, X.~Cao, C.~Cai, \emph{et~al.}, \enquote{Multidimensional fiber-to-chip optical processing using photonic integrated circuits,} {\protect\JournalTitle{Laser \& Photonics Reviews}} \textbf{18}, 2300853 (2024).

\bibitem{montaut2025progress}
N.~Montaut, A.~George, M.~Monika, \emph{et~al.}, \enquote{Progress in integrated and fiber optics for time-bin based quantum information processing,} {\protect\JournalTitle{Advanced Optical Technologies}} \textbf{14}, 1560084 (2025).

\bibitem{myilswamy2024on-chip}
K.~V. Myilswamy, L.~M. Cohen, S.~Seshadri, \emph{et~al.}, \enquote{On-chip frequency-bin quantum photonics,} {\protect\JournalTitle{Nanophotonics}} pp. 1--16 (2024).

\bibitem{corrielli2021femtosecond}
G.~Corrielli, A.~Crespi, and R.~Osellame, \enquote{Femtosecond laser micromachining for integrated quantum photonics,} {\protect\JournalTitle{Nanophotonics}} \textbf{10}, 3789--3812 (2021).

\bibitem{li2018multimode}
C.~Li, D.~Liu, and D.~Dai, \enquote{Multimode silicon photonics,} {\protect\JournalTitle{Nanophotonics}} \textbf{8}, 227--247 (2018).

\bibitem{cristiani2022roadmap}
\enquote{Roadmap on multimode photonics,} {\protect\JournalTitle{Journal of Optics}} \textbf{24}, 083001 (2022).

\bibitem{yang2022multi}
K.~Y. Yang, C.~Shirpurkar, A.~D. White, \emph{et~al.}, \enquote{Multi-dimensional data transmission using inverse-designed silicon photonics and microcombs,} {\protect\JournalTitle{Nature communications}} \textbf{13}, 7862 (2022).

\bibitem{sun2025edge}
A.~Sun, S.~Xing, X.~Deng, \emph{et~al.}, \enquote{Edge-guided inverse design of digital metamaterial-based mode multiplexers for high-capacity multi-dimensional optical interconnect,} {\protect\JournalTitle{Nature Communications 2025 16:1}} \textbf{16}, 1--12 (2025).

\bibitem{jiang2019compact}
W.~Jiang, J.~Miao, and T.~Li, \enquote{Compact silicon 10-mode multi/demultiplexer for hybrid mode-and polarisation-division multiplexing system,} {\protect\JournalTitle{Scientific reports}} \textbf{9}, 13223 (2019).

\bibitem{xu2018metamaterial2mode}
H.~Xu and Y.~Shi, \enquote{Metamaterial-based maxwell's fisheye lens for multimode waveguide crossing,} {\protect\JournalTitle{Laser \& Photonics Reviews}} \textbf{12}, 1800094 (2018).

\bibitem{guo2023transmitarray3mode}
X.~Guo, Z.~Wang, Y.~Zhang, \emph{et~al.}, \enquote{Ultra-broadband multimode waveguide crossing via subwavelength transmitarray with bound state,} {\protect\JournalTitle{Laser \& Photonics Reviews}} \textbf{17}, 2200674 (2023).

\bibitem{jiang2018low-loss}
\enquote{Low-loss and low-crosstalk multimode waveguide bend on silicon,} {\protect\JournalTitle{Optics Express, Vol. 26, Issue 13, pp. 17680-17689}} \textbf{26}, 17680--17689 (2018).

\bibitem{priti2019reconfigurable}
R.~B. Priti and O.~Liboiron-Ladouceur, \enquote{Reconfigurable and scalable multimode silicon photonics switch for energy-efficient mode-division-multiplexing systems,} {\protect\JournalTitle{Journal of Lightwave Technology}} \textbf{37}, 3851--3860 (2019).

\bibitem{feng2016chip}
L.-T. Feng, M.~Zhang, Z.-Y. Zhou, \emph{et~al.}, \enquote{On-chip coherent conversion of photonic quantum entanglement between different degrees of freedom,} {\protect\JournalTitle{Nature communications}} \textbf{7}, 11985 (2016).

\bibitem{feng2019chip}
L.-T. Feng, M.~Zhang, X.~Xiong, \emph{et~al.}, \enquote{On-chip transverse-mode entangled photon pair source,} {\protect\JournalTitle{npj Quantum Information}} \textbf{5}, 2 (2019).

\bibitem{hong1987measurement}
C.-K. Hong, Z.-Y. Ou, and L.~Mandel, \enquote{Measurement of subpicosecond time intervals between two photons by interference,} {\protect\JournalTitle{Physical review letters}} \textbf{59}, 2044 (1987).

\bibitem{mohanty2017quantum}
A.~Mohanty, M.~Zhang, A.~Dutt, \emph{et~al.}, \enquote{Quantum interference between transverse spatial waveguide modes,} {\protect\JournalTitle{Nature communications}} \textbf{8}, 1--7 (2017).

\bibitem{feng2022transverse}
L.-T. Feng, M.~Zhang, X.~Xiong, \emph{et~al.}, \enquote{Transverse mode-encoded quantum gate on a silicon photonic chip,} {\protect\JournalTitle{Physical Review Letters}} \textbf{128}, 060501 (2022).

\bibitem{Su2020}
L.~Su, D.~Vercruysse, J.~Skarda, \emph{et~al.}, \enquote{Nanophotonic inverse design with spins: Software architecture and practical considerations,} {\protect\JournalTitle{Applied Physics Reviews}} \textbf{7}, 11407 (2020).

\bibitem{Lalau-Keraly2013}
\enquote{Adjoint shape optimization applied to electromagnetic design,} {\protect\JournalTitle{Optics Express, Vol. 21, Issue 18, pp. 21693-21701}} \textbf{21}, 21693--21701 (2013).

\bibitem{Callewaert2016}
F.~Callewaert, S.~Butun, Z.~Li, and K.~Aydin, \enquote{Inverse design of an ultra-compact broadband optical diode based on asymmetric spatial mode conversion,} {\protect\JournalTitle{Scientific Reports 2016 6:1}} \textbf{6}, 1--10 (2016).

\bibitem{hansen2024inverse}
S.~E. Hansen, G.~Arregui, A.~N. Babar, \emph{et~al.}, \enquote{Inverse design and characterization of compact, broadband, and low-loss chip-scale photonic power splitters,} {\protect\JournalTitle{Materials for Quantum Technology}} \textbf{4}, 016201 (2024).

\bibitem{shang2023inverse-designed}
C.~Shang, J.~Yang, A.~M. Hammond, \emph{et~al.}, \enquote{Inverse-designed lithium niobate nanophotonics,} {\protect\JournalTitle{ACS Photonics}} \textbf{10}, 1019--1026 (2023).

\bibitem{jiang2023inverse}
W.~Jiang, S.~Mao, and J.~Hu, \enquote{Inverse-designed counter-tapered coupler based broadband and compact silicon mode multiplexer/demultiplexer,} {\protect\JournalTitle{Optics Express}} \textbf{31}, 33253--33263 (2023).

\bibitem{zhang2021scalable}
G.~Zhang and O.~Liboiron-Ladouceur, \enquote{Scalable and low crosstalk silicon mode exchanger for mode division multiplexing system enabled by inverse design,} {\protect\JournalTitle{IEEE Photonics Journal}} \textbf{13}, 1--13 (2021).

\bibitem{Shen2024}
R.~Shen, R.~Shen, F.~Hu, \emph{et~al.}, \enquote{100nm broadband and ultra-compact multi-dimensional multiplexed photonic integrated circuit for high-capacity optical interconnects,} {\protect\JournalTitle{Photonics Research, Vol. 12, Issue 12, pp. 2891-2900}} \textbf{12}, 2891--2900 (2024).

\bibitem{yang2022inverse}
S.~Yang, H.~Jia, J.~Niu, \emph{et~al.}, \enquote{Cmos-compatible ultra-compact silicon multimode waveguide bend based on inverse design method,} {\protect\JournalTitle{Optics Communications}} \textbf{523}, 128733 (2022).

\bibitem{he2023super}
L.~He, D.~Liu, J.~Gao, \emph{et~al.}, \enquote{Super-compact universal quantum logic gates with inverse-designed elements,} {\protect\JournalTitle{Science Advances}} \textbf{9}, eadg6685 (2023).

\bibitem{huang2025multiphoton}
S.-Y. Huang, S.~Kumar, J.~Huster, \emph{et~al.}, \enquote{Multiphoton quantum interference at ultracompact inverse-designed multiport beam splitter,} {\protect\JournalTitle{arXiv preprint arXiv:2504.00114}}  (2025).

\bibitem{bouchard2020two}
F.~Bouchard, A.~Sit, Y.~Zhang, \emph{et~al.}, \enquote{Two-photon interference: the hong--ou--mandel effect,} {\protect\JournalTitle{Reports on Progress in Physics}} \textbf{84}, 012402 (2020).

\bibitem{fedrizzi2009anti}
A.~Fedrizzi, T.~Herbst, M.~Aspelmeyer, \emph{et~al.}, \enquote{Anti-symmetrization reveals hidden entanglement,} {\protect\JournalTitle{New Journal of Physics}} \textbf{11}, 103052 (2009).

\bibitem{hughes2018adjoint}
T.~W. Hughes, M.~Minkov, I.~A. Williamson, and S.~Fan, \enquote{Adjoint method and inverse design for nonlinear nanophotonic devices,} {\protect\JournalTitle{ACS Photonics}} \textbf{5}, 4781--4787 (2018).

\bibitem{Nanda2024ExploringOptimization}
A.~Nanda, M.~Kues, and A.~Cal{\`{a}}~Lesina, \enquote{{Exploring the fundamental limits of integrated beam splitters with arbitrary phase via topology optimization},} {\protect\JournalTitle{Optics Letters}} \textbf{49}, 1125 (2024).

\bibitem{Vercruysse2019}
D.~Vercruysse, N.~V. Sapra, L.~Su, \emph{et~al.}, \enquote{Analytical level set fabrication constraints for inverse design,} {\protect\JournalTitle{Scientific Reports 2019 9:1}} \textbf{9}, 1--7 (2019).

\bibitem{ghafari2025entanglement}
F.~Ghafari, J.~Dias, L.~K. Shalm, \emph{et~al.}, \enquote{Entanglement distillation rates exceeding the direct transmission bound,} {\protect\JournalTitle{arXiv preprint arXiv:2503.21133}}  (2025).

\bibitem{White2025timebin}
S.~J.~U. White, E.~Polino, F.~Ghafari, \emph{et~al.}, \enquote{Robust approach for time-bin-encoded photonic quantum information protocols,} {\protect\JournalTitle{Phys. Rev. Lett.}} \textbf{134}, 180802 (2025).

\bibitem{wang2020qudits}
Y.~Wang, Z.~Hu, B.~C. Sanders, and S.~Kais, \enquote{Qudits and high-dimensional quantum computing,} {\protect\JournalTitle{Frontiers in Physics}} \textbf{8}, 589504 (2020).

\bibitem{kiktenko2023realization}
E.~O. Kiktenko, A.~S. Nikolaeva, and A.~K. Fedorov, \enquote{Realization of quantum algorithms with qudits,} {\protect\JournalTitle{arXiv preprint arXiv:2311.12003}}  (2023).

\bibitem{de2018simple}
H.~de~Guise, O.~Di~Matteo, and L.~L. S{\'a}nchez-Soto, \enquote{Simple factorization of unitary transformations,} {\protect\JournalTitle{Physical Review A}} \textbf{97}, 022328 (2018).

\bibitem{saygin2020robust}
M.~Y. Saygin, I.~Kondratyev, I.~Dyakonov, \emph{et~al.}, \enquote{Robust architecture for programmable universal unitaries,} {\protect\JournalTitle{Physical review letters}} \textbf{124}, 010501 (2020).

\bibitem{reck1994experimental}
M.~Reck, A.~Zeilinger, H.~J. Bernstein, and P.~Bertani, \enquote{Experimental realization of any discrete unitary operator,} {\protect\JournalTitle{Physical review letters}} \textbf{73}, 58 (1994).

\bibitem{clements2016optimal}
W.~R. Clements, P.~C. Humphreys, B.~J. Metcalf, \emph{et~al.}, \enquote{Optimal design for universal multiport interferometers,} {\protect\JournalTitle{Optica}} \textbf{3}, 1460--1465 (2016).

\end{thebibliography}
\end{document}